# On noise-resolution uncertainty in quantum field theory

Timur E. Gureyev,[1,2,3,4,*] Alexander Kozlov,[1] Yakov I. Nesterets,[4,3] David M. Paganin,[2] and Harry M. Quiney[1]

[1] *ARC Centre in Advanced Molecular Imaging, School of Physics, The University of Melbourne, Parkville 3010, Australia*

[2] *School of Physics and Astronomy, Monash University, Clayton 3800, Australia*

[3] *School of Science and Technology, University of New England, Armidale 2351, Australia*

[4] *CSIRO Manufacturing, Clayton 3168, Australia*

*Correspondence and requests for materials should be addressed to T.E.G. (email: timur.gureyev@unimelb.edu.au)*

**Abstract**

An uncertainty inequality is presented that establishes a lower limit for the product of the variance of the time-averaged intensity of a mode of a quantized electromagnetic field and the degree of its spatial localization. The lower limit is determined by the vacuum fluctuations within the volume corresponding to the width of the mode. This result also leads to a generalized form of the Heisenberg uncertainty principle for boson fields in which the lower limit for the product of uncertainties in the spatial and momentum localization of a mode is equal to the product of Planck's constant and a dimensionless functional which reflects the joint signal-to-noise ratio of the position and momentum of vacuum fluctuations in the region of the phase space occupied by the mode. Experimental X-ray synchrotron measurements provide an initial verification of the proposed theory in the case of Poisson statistics.

**Introduction**

A cornerstone of quantum mechanics is the fact that certain quantities, like position and momentum of a particle, cannot be measured simultaneously with arbitrary precision. Similarly, quantum field theory sets a limit for the precision of certain measurements due to the presence of vacuum fluctuations. The two phenomena are not equivalent, but they are related. In the present work we establish a quantitative relationship between these two aspects of quantum measurement, which has the form of a noise-resolution uncertainty inequality involving the variance of the position or momentum of the field's mode on one hand and the variance of the energy of the mode on the other. We show that the product of the two variances can never be smaller than a certain absolute positive constant. Importantly, the uncertainty relationship derived here is qualitatively different from the classical Heisenberg uncertainty [1-3], as it reaches its minimum not for Gaussian distributions or coherent states, but for the so-called Epanechnikov distributions that are well known in mathematical statistics [4-6]. This particular fact distinguishes our uncertainty relationship from the number-phase uncertainty [7]. There are several other known uncertainty relationships related to the Heisenberg uncertainty [8-15]. In particular, the Robertson uncertainty principle [8] represents a generalized form of the Heisenberg uncertainty. In turn, the Robertson uncertainty follows directly from the Schrödinger uncertainty inequality [9]. As far as the physical meaning of the Heisenberg uncertainty is concerned, different interpretations have been discussed, starting from Heisenberg's own interpretation [1] and the similar interpretation by Feynman [2]. In particular, the difference between the disturbance to the state of the system produced by the measurement ("systematic errors") and the variance in the measured quantities as a result of "statistical imprecision" of the measurements have been analysed as different sources contributing to the overall uncertainty in the measurements [10, 12]. Reviews of the history of the subject and related concepts can be found in [3, 10, 14].

As a motivating example for this Report, consider a three-dimensional volume of space which is permeated by a quantised electromagnetic field. Suppose that the number of photons which will be extracted from this field is a specified integer, and that the volume over which photon detection is to take place is also fixed. In building up an intensity map of localisable features in the photon field, there is an evident practical trade-off between noise and resolution; increasing the number of spatial resolution elements ("voxels") in the fixed specified volume will increase resolution (i.e. decrease voxel size) of the detected intensity signal, at the expense of increased relative noise in the intensity signal. The first key result of this Report is a noise-resolution uncertainty principle which quantifies this trade-off by establishing a lower bound on the product of the spatial resolution volume and the variance of the intensity signal. Note that localization of a field can be achieved either by the use of a detector with sufficient spatial resolution or by considering fields which contain a single spatially-localized feature. For the purposes of our study, the two approaches are essentially equivalent and in the present Report we adopt the latter.

The noise-resolution uncertainty principle reported here is neither reducible to, nor derivable from Heisenberg´s uncertainty principle. Moreover, when applied to both the uncertainty of the momentum and position of a boson field, it leads to a modified form of the Heisenberg uncertainty inequality in which the reduced Planck constant on the right-hand side is multiplied by a dimensionless functional reflecting the signal-to-noise ratio (SNR) of the momentum and position of vacuum fluctuations in the region of the phase space occupied by the mode. When the value of this functional is equal to unity, the modified inequality reduces to the conventional Heisenberg uncertainty inequality. However, when the value of the functional is large, our modified inequality predicts that the minimum of the product of uncertainties of the momentum and position of the field can actually be larger than the one predicted by the Heisenberg uncertainty. The additional uncertainty is associated with the trade-off between the precision of a simultaneous measurement of position and momentum, and the SNR of these measurements.

The trade-off between spatial resolution and SNR is important for many imaging and communication problems, where simultaneous optimization of both spatial resolution and SNR is usually desired [16, 17]. A similar trade-off between noise and spatial resolution can in fact be demonstrated in the case of some fundamental experiments involving quantum measurements, such as Young's double-slit experiment with electrons [2, 18, 19].

**Results**

*Noise-resolution uncertainty inequality for boson fields*

Consider a linearly-polarized quantised electromagnetic field defined by an operator-valued distribution $E(\mathbf{x}) = \sum_k E_k(\mathbf{x})$, where

$E_k(\mathbf{x}) = i(\hbar\omega_k/2)^{1/2}[u_k(\mathbf{r})\exp(-i\omega_k t)a_k - u_k^*(\mathbf{r})\exp(i\omega_k t)a_k^\dagger]$, $\mathbf{x} = (\mathbf{r}, t)$ is a four-dimensional space-time point, $\hbar$ is the reduced Planck constant, $\omega_k$ are the angular frequencies of the modes, $u_k(\mathbf{r})$ are the mode functions, $a_k$ and $a_k^\dagger$ are the photon annihilation and creation operators, respectively. We assume the mode functions $u_k(\mathbf{r})$ to be orthonormal, in the sense that $\int d\mathbf{r}\, u_{k_1}^*(\mathbf{r}) u_{k_2}(\mathbf{r}) = \delta_{k_1 k_2}$, where $\delta_{k_1 k_2}$ is the Kronecker symbol, but we do not limit the choice of these functions only to (truncated) plane waves. We study the behaviour of the time-averaged space-integrated variance of the field's intensity, $(1/2)E^2(\mathbf{r}, t)$:

$$(\Delta I_\psi)^2 \equiv \lim_{T\to\infty} \frac{1}{T} \int_{-T/2}^{T/2} dt \int d\mathbf{r}(1/4)[<\psi|E^4(\mathbf{r},t)|\psi> - (<\psi|E^2(\mathbf{r},t)|\psi>)^2]. \qquad (1)$$

We are primarily interested in the states of the field with small numbers of particles, where the effect of vacuum fluctuations can be prominent. In such cases it is often convenient to work with the Fock space representation. Using the standard properties of the Fock states, $|n>$, and the field creation and annihilation operators [7], it can be shown by direct calculation that, for a single-mode case, one has:

$$<n|E_k^2(\mathbf{x})|n> = \hbar\omega_k(n+1/2)|u_k(\mathbf{r})|^2,$$
$$<n|E_k^4(\mathbf{x})|n> = (3/2)(\hbar\omega_k)^2[(n+1/2)^2 + 1/4]|u_k(\mathbf{r})|^4,$$

and hence,

$$(\Delta I_n)^2 = (1/4)\int d\mathbf{r}[<n|E_k^4(\mathbf{x})|n> - (<n|E_k^2(\mathbf{x})|n>)^2]$$
$$= [(n+1/2)^2 + 3/4][(\hbar\omega_k)^2/8]\int d\mathbf{r}|u_k(\mathbf{r})|^4 \geq (\Delta I_0)^2, \qquad (2)$$

where $(\Delta I_0)^2 = [(\hbar\omega_k)^2/8]\int d\mathbf{r}|u(\mathbf{r})|^4$ is the intensity variance of the vacuum state $|0>$. Using a conventional variational approach, it is also possible to show that the variance of field intensity in any state cannot be smaller than that of a vacuum state. As an example, explicit calculations for the intensity variance in a (Glauber) coherent state $|\alpha>$ [7, 20] produce:

$$(\Delta I_\alpha)^2 = \lim_{T\to\infty} \frac{1}{T} \int_{-T/2}^{T/2} dt \int d\mathbf{r}(1/4)[<\alpha|E_k^4(\mathbf{x})|\alpha> - (<\alpha|E_k^2(\mathbf{x})|\alpha>)^2]$$
$$= (|\alpha|^2 + 1/2)[(\hbar\omega_k)^2/4]\int d\mathbf{r}|u(\mathbf{r})|^4 \geq (\Delta I_0)^2. \qquad (3)$$

Notice that, because the mode functions are normalized in the sense that $\int d\mathbf{r}\,|u_k(\mathbf{r})|^2 = 1$, a narrow (tightly localized) mode function will correspond to large intensity variance in equations (2)-(3). This behaviour can be easily seen in the example of a truncated plane wave, $u_k(\mathbf{r}) = L_k^{-3/2}\exp(i\mathbf{k}\cdot\mathbf{r})\chi_{L_k}(\mathbf{r})$, where $\chi_{L_k}(\mathbf{r})$ is equal to one inside a cube $V_{L_k}$ centred at $\mathbf{r} = 0$, with sides of length $L_k$ parallel to the coordinate axes, and $\chi_{L_k}(\mathbf{r})$ is equal to zero outside the cube. Here $\int d\mathbf{r}\,|u_k(\mathbf{r})|^2 = L_k^{-3}\int d\mathbf{r}\,\chi_{L_k}(\mathbf{r}) = 1$ and $(\Delta I)^2 \propto \int d\mathbf{r}\,|u_k(\mathbf{r})|^4 = L_k^{-3} \xrightarrow[L_k \to 0]{} \infty$. The observation that the variance of the field intensity increases with increased spatial localization is made quantitative by the "noise-resolution uncertainty principle" described below.

We define the spatial width of a mode as

$$(\Delta r)^2 = \int d\mathbf{r}\,|\mathbf{r} - \bar{\mathbf{r}}|^2\,|u_k(\mathbf{r})|^2, \tag{4}$$

where $\bar{\mathbf{r}}$ is the mean value of variable $\mathbf{r}$ with probability density function $|u_k(\mathbf{r})|^2$. It was proven in [6] that for any positive integer dimension $d$ and any function $f \in L_1(\mathbf{R}^d) \cap L_1(\mathbf{R}^d, |\mathbf{y}|^2) \cap L_2(\mathbf{R}^d)$, the following uncertainty inequality holds:

$$\frac{\|f\|_2^4\,\left\||\mathbf{y}-\mathbf{y}_0|^2 f\right\|_1^d}{\|f\|_1^{d+4}} \geq \left(\frac{d}{4\pi}\right)^d (C_d)^2,\quad C_d = 2^d\,\Gamma(d/2)\,d(d+2)/(d+4)^{d/2+1}, \tag{5}$$

where $\|f\|_p = \left(\int d\mathbf{y}\,|f(\mathbf{y})|^p\right)^{1/p}$. The equality in equation (5) is achieved for Epanechnikov distributions $f_E(\mathbf{y}) = c_1(1 - |c_2\mathbf{y} - \mathbf{y}_0|^2)_+$, where $c_1$ and $c_2$ are arbitrary positive constants, and the subscript "+" denotes that the function is equal to zero for those values of its argument, where the expression inside the brackets is negative [6]. Taking $d = 3$ and $f(\mathbf{r}) = |u(\mathbf{r})|^2$, we obtain from equations (2)-(5) the first key result of this Report:

$$(\Delta r)^3 (\Delta I_\psi)^2 / W_0^2 \geq 2 \int d\mathbf{r} \, |u_k(\mathbf{r})|^4 \left( \int d\mathbf{r} \, |\mathbf{r} - \overline{\mathbf{r}}|^2 |u_k(\mathbf{r})|^2 \right)^{3/2} \geq \tilde{C}_3, \tag{6}$$

where $W_0 = \hbar \omega_k / 4$ is the energy of the vacuum state of the electric field and $\tilde{C}_3 = (3/\pi)^{3/2} C_3 / 4 = 15\sqrt{27} / (7^{5/2} \pi) \cong 0.19$ is a dimensionless constant. This inequality expresses an uncertainty relationship that exists between the noise in the measured intensity of a field mode and the variance of its spatial distribution which quantifies the degree of its spatial localization. Using equation (5), the noise-resolution uncertainty relationship can be easily generalized to arbitrary (integer) dimension $d = 1,2,3,...$. The case $d = 2$, for example, can be applied to cross-sections of beams in planes orthogonal to the optical axis.

It is possible to re-formulate equation (6) in terms of the spatial averages of the intensity, $\tilde{I}_\psi = \lim_{T \to \infty} \frac{1}{T} \int_{-T/2}^{T/2} dt \frac{1}{L^3} \int_{V_L} d\mathbf{r} (1/2) < \psi | E_k^2(\mathbf{x}) | \psi >$, and the intensity variance,

$(\Delta \tilde{I}_\psi)^2 = \lim_{T \to \infty} \frac{1}{T} \int_{-T/2}^{T/2} dt \frac{1}{L^3} \int_{V_L} d\mathbf{r} (1/4) [< \psi | E_k^4(\mathbf{x}) | \psi > - (< \psi | E_k^2(\mathbf{x}) | \psi >)^2]$, over a cube $V_L$. In this case the relative noise, $(\Delta \tilde{I}_\psi)^2 / \tilde{I}_0^2$, is dimensionless, while the spatial resolution volume $(\Delta r)^3$ is replaced by its relative value, $(\Delta r)^3 / L^3$, which can be interpreted as the fraction of the cube's volume occupied by the mode. As a result, an analogue of equation (6) can be written as:

$$\frac{(\Delta r)^3}{L^3} \frac{(\Delta \tilde{I}_\psi)^2}{\tilde{I}_0^2} \geq \tilde{C}_3. \tag{7}$$

The limit of this inequality at $L \to \infty$ coincides with equation (6).

To appreciate the behaviour of equation (6), consider the case of a truncated plane wave, $u_k(\mathbf{r}) = L_k^{-3/2} \exp(i\mathbf{k} \cdot \mathbf{r}) \chi_{L_k}(\mathbf{r})$, in a Fock state $|n>$ or in a coherent state $|\alpha>$. From equations (2)-(4) we obtain: $(\Delta r)^3 = (L_k/2)^3$, $(\Delta I_n)^2 / W_0^2 = 2[(n+1/2)^2 + 3/4] L_k^{-3}$ and $(\Delta I_\alpha)^2 / W_0^2 = 4(|\alpha|^2 + 1/2) L_k^{-3}$. Clearly, here the values of the spatial resolution and the relative noise are counter-balanced with respect to the parameter $L_k$, with the expression in equation (6) being equal to $(\Delta r)^3 (\Delta I_n)^2 / W_0^2 = [(n+1/2)^2 + 3/4]/4 \geq 1/4 > \tilde{C}_3$ and $(\Delta r)^3 (\Delta I_\alpha)^2 / W_0^2 = (|\alpha|^2 + 1/2)/2 \geq 1/4 > \tilde{C}_3$, respectively. One can also see that when the photon energy statistics are known, inequalities (6) and (7) can sometimes be considerably strengthened. For example, if the photon energy statistics are Gaussian, as in the case of Fock states, then it can be seen that the ratio $(\Delta I_n)^2 / W_n^2 = (1/2)[1 + 3(2n+1)^{-2}] \int d\mathbf{r} |u_k(\mathbf{r})|^4$, where $W_n = W_0 (2n+1)$ is the energy of the mode in state $|n>$, quickly approaches a constant asymptote when $n$ increases. As a result, equation (6) can be replaced by a stronger inequality: $(\Delta r)^3 (\Delta I_n)^2 / W_n^2 \geq \tilde{C}_3 / 4$. If the photon statistics are Poissonian, as in the case of coherent states, then $(\Delta r)^3 (\Delta I_\alpha)^2 / W_\alpha^2 \geq \tilde{C}_3 / (2\bar{n} + 1)$, where $W_\alpha = W_0 (2|\alpha|^2 + 1)$. Consequently, equations (6)-(7) can be modified in the general $d$-dimensional case to give the following noise-resolution uncertainty principle:

$$\lim_{L\to\infty} \frac{1}{M_L SNR_{\psi,L}^2} \equiv \lim_{L\to\infty} \frac{(\Delta r)^d}{L^d} \frac{(\Delta \tilde{I}_\psi)^2}{\tilde{I}_\psi^2} = (\Delta r)^d \frac{(\Delta I_\psi)^2}{W_\psi^2} \geq \frac{\tilde{C}_{d,\gamma}}{(\bar{n} + 1/2)^\gamma}, \quad (8)$$

where $M_L = L^d / (\Delta r)^d$ is the number of spatial resolution units in the measuring system, $SNR_{\psi,L} = \tilde{I}_\psi / \Delta \tilde{I}_\psi$ is the average SNR, $\tilde{C}_{d,\gamma}$ is a positive constant which depends only on the dimensionality of the space and the photon energy statistics, and $\gamma = 0, 1$ or $2$, respectively, in the case of Gaussian statistics, Poissonian statistics and generic statistics (which corresponds to equation (6) and is valid in all cases, including sub-Poissonian statistics, in particular). In the case of Poissonian statistics ($\gamma = 1$), equation (8) is close in form to the noise-resolution uncertainty previously demonstrated in the context of X-ray imaging [17, 21]. It has been shown in [21] that the quantity $Q_d^2 = [d/(4\pi)]^{d/2} M\, SNR^2 / (\bar{n} + 1/2)$ has characteristics somewhat similar to "information capacity per single particle". Equation (8) is equivalent to the statement that $Q_d^2$ cannot exceed an absolute upper limit: $Q_d^2 \leq 1/C_d$.

*Experimental test of the noise-resolution uncertainty relation*

We have carried out an initial experimental verification of equation (8) using data collected at the Imaging and Medical beamline of the Australian Synchrotron. The essential feature of this result presented in Fig.1 is the observed approximately constant behaviour of $Q_2^2(M,\bar{n})$ (see details in the Methods section) within the tested range of parameters, i.e. its invariance with respect to the spatial resolution and the number of photons. According to the above theory in the case of Poissonian statistics, i.e. when $SNR^2$ is proportional to $\bar{n}$, one should get $Q_2^2(M,\bar{n}) = 1/C[f]$, where $C[f] = 2\pi \int |\mathbf{r}_\perp - \bar{\mathbf{r}}_\perp|^2 f(\mathbf{r}_\perp)d\mathbf{r}_\perp \int f^2(\mathbf{r}_\perp)d\mathbf{r}_\perp$ is a positive constant which depends only on the shape of the point-spread function, $f(\mathbf{r}_\perp) \equiv |u_k(\mathbf{r}_\perp)|^2$, of the detector, with $\mathbf{r}_\perp$ being a two-dimensional coordinate on the entrance surface of the detector. Therefore, the behaviour of the experimental data shown in Fig.1 is in agreement with equation (8) for $d = 2$. Note that, given the observed independence of $Q_2^2(M,\bar{n})$ from $M$ and $\bar{n}$, the existence of a positive absolute lower limit in equation (8) is then a simple consequence of the mathematical inequality (5) which holds for any function $f(\mathbf{r}_\perp)$. Much more sophisticated experiments may be required in order to verify equations (6)-(8) at very low photon levels, where the effect of quantum fluctuations can be detected directly. Such an experiment may be possible using techniques similar to those employed recently in single-particle diffraction experiments [18] or with femtosecond laser pulses [22].

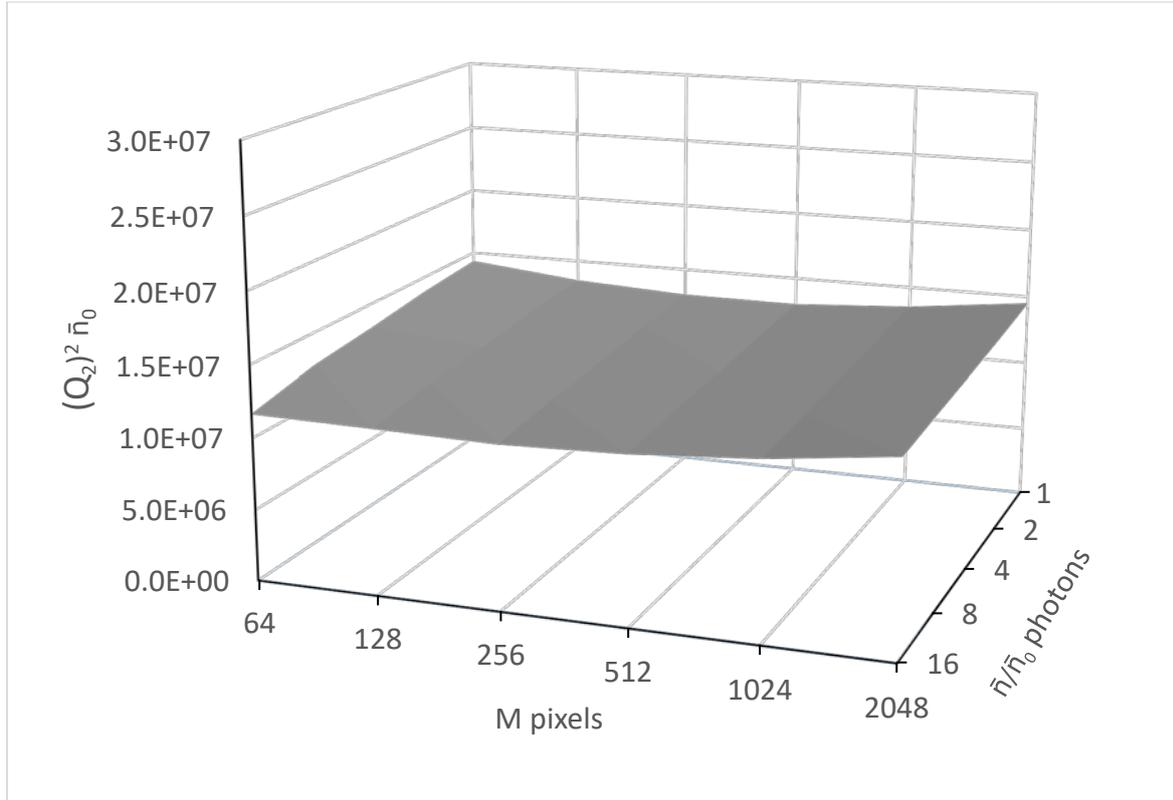

Fig.1. Measured dependence of the quantity $Q_2^2$ on the spatial resolution (number of effective pixels *M*) and on the total number of photons.

*An extension of the Heisenberg uncertainty relation*

Using the same approach, as in the derivation of equation (6), but for the Fourier transform (momentum representation) of the mode, we obtain:

$$(\Delta \xi)^3 (\Delta \tilde{I}_\psi)^2 / W_0^2 = 2 \int d\xi \, |\tilde{u}_k(\xi)|^4 \left( \int d\xi \, |\xi - \overline{\xi}|^2 |\tilde{u}_k(\xi)|^2 \right)^{3/2} \geq \tilde{C}_3, \qquad (9)$$

where $\tilde{u}_k(\xi) = \int d\mathbf{r} \exp(-i2\pi \xi \cdot \mathbf{r}) u_k(\mathbf{r})$. Multiplying equations (6) and (9), and taking into account the Heisenberg uncertainty inequality for $d = 3$ [11], $\Delta r \Delta p \geq 3\hbar / 2$, we obtain the second key result of this Report - an extension of the Heisenberg uncertainty principle:

$$\Delta r \Delta p \geq (3\hbar/2) \max\{1, SNR_0[u_k]\}, \tag{10}$$

where $\Delta p = 2\pi\hbar\Delta\xi$, and

$$SNR_0[u_k] \equiv \left(\frac{4C_3^2 W_0^4}{(\Delta I_0)^2(\Delta \tilde{I}_0)^2}\right)^{1/3} = \frac{C_3^{2/3}\left(\int d\mathbf{r}\,|u_k(\mathbf{r})|^2\right)^{4/3}}{\left(\int d\mathbf{r}\,|u_k(\mathbf{r})|^4 \int d\boldsymbol{\xi}\,|\tilde{u}_k(\boldsymbol{\xi})|^4\right)^{1/3}}, \tag{11}$$

is a non-negative dimensionless functional defined on modes $u_k(\mathbf{r})$. We included the normalization of the mode to unity into the numerator of equation (11) to make more obvious the fact that the functional $SNR_0[u_k]$ is bi-invariant with respect to multiplication of the mode or its argument **r** by an arbitrary positive number. That means that the value of this functional does not depend on the scaling of the height and width of the function on which it acts, but only on the functional form. Unlike the similar bi-invariant positive functionals appearing in the Heisenberg uncertainty [11] and in equation (5), however, it can be shown that the value of $SNR_0[u_k]$ can be arbitrarily large or arbitrarily close to zero for some functions, $u_k(\mathbf{r})$ [21]. When $SNR_0[u_k] < 1$, the inequality $\Delta r \Delta p \geq (3\hbar/2)SNR_0[u_k]$ is weaker than the Heisenberg uncertainty inequality, and hence equation (10) does not show any new effects in this instance. However, when $SNR_0[u_k] > 1$, inequality (10) implies that the product of uncertainties in the position and momentum of the mode has a lower limit that is larger than the one given by the Heisenberg uncertainty. In other words, inequality (10) is stronger in this case than the Heisenberg uncertainty, giving what we term an anti-squeezed Heisenberg uncertainty principle. For example, for a plane-wave mode in a cube with side length $L_k$, $u_k(\mathbf{r}) = L_k^{-3/2}\exp(i\mathbf{k}\cdot\mathbf{r})\chi_{L_k}(\mathbf{r})$, we obtain $SNR_0[u_k] = (3/2)C_3^{2/3} \cong 1.3$ regardless of the size of the cube. For such functions, inequality (10) already gives a larger lower bound for the minimal phase-space volume than the Heisenberg uncertainty.

**Discussion**

We have shown in equation (6) that the product of the variance of electric energy of a mode and the variance of its spatial distribution, cannot be made smaller than the square of the energy of the mode in the vacuum state multiplied by a certain positive dimensionless constant. In this context, the minimum uncertainty state is achieved for Epanechnikov distributions [6]. We then demonstrated in equation (8) that if the photon energy statistics is known, the uncertainty relation between the precision of spatial localization of the mode and the precision of its intensity measurement can be specialized and made stronger for different types of the statistics. It could be interesting to consider the possibility of extending the noise-resolution uncertainty similar to equation (6) and equation (8) from the energy of an electromagnetic mode to the photon number operator. One could argue that in most optical experiments it is the number of photons, rather than the field energy, that is measured by the detector [7, 20] (note, however, the result of recent direct measurements of vacuum energy fluctuations [22]). In relation to the noise-resolution uncertainty, the key issue is that while the variance of field's energy has a positive absolute lower limit, due to the existence of vacuum fluctuations, the variance of photon number operator can be zero, as e.g. in Fock states. Therefore, direct analogues of inequalities (2)-(3) cannot exist for photon number operator. Nevertheless, it is also known that states with low photon number variance have weak spatial localization. Therefore, a noise-resolution uncertainty relation between the width of spatial distribution of a mode and the variance of the number of photons can still exist, but a relevant proof will have to be different from the one used for equation (6) above.

Now let us further consider the physical meaning of the second main result of this paper, i.e. an extension of the Heisenberg uncertainty in the form of equation (10). It is evident from the definition of $SNR_0[u_k]$ in equation (11), that the case $SNR_0[u_k] > 1$ generally corresponds to measurements with low variance or high SNR. This result may seem counter-intuitive at first, as it states that joint high-SNR measurements of the position and momentum of an electromagnetic field must produce a very imprecise result in terms of the position and momentum localization. This is, however, a direct consequence of the noise-resolution uncertainty principle which makes the position and momentum variance larger, whenever the noise level in the corresponding measurement decreases towards its lower limit. Therefore, equation (10) can be viewed as a logical generalization of the Heisenberg uncertainty to the case of measurements involving multiple particles (boson fields). In this case, the uncertainty in measurements of conjugate observables can be traded not only for each other, but also for the SNR in the measurement of each observable. The "non-reducible" quantity here is not the minimal phase-space volume $\Delta r \Delta p$, but the ratio of $\Delta r \Delta p$ and the "joint vacuum SNR" of the measurements, $\max\{1, SNR_0[u_k]\}$. On the other hand, when this SNR becomes less than 1, the classical Heisenberg uncertainty takes over, with the consequence that no improvement in the accuracy of simultaneous measurements of the position and momentum beyond the Heisenberg limit can be achieved using measurements with low SNR. Of course, these conclusions have to be viewed in the context of the specific meaning of the "joint vacuum SNR" as defined by equation (11).

**Methods**

In the Results section above, we used simple direct calculations for the energy of the electromagnetic field in the coherent and Fock states. One could argue that more general calculations could be attempted using the formalism of *n*-point correlation functions often employed for similar purposes in quantum optics. It turns out that, for a field state with a low number of photons, the latter approach, if used without attention to relevant inherent approximations, can lead to incorrect results. It is instructive to consider this methodological issue here as it provides an insight into the nature of our central result, equation (6). Let the first and the second order correlation functions of the field $E(\mathbf{x})$ have the usual form [7, 20]:

$$G^{(1)}(\mathbf{x}_1, \mathbf{x}_2) = Tr(\rho E^{(-)}(\mathbf{x}_1) E^{(+)}(\mathbf{x}_2))$$
$$= \sum_{k_1, k_2} w^*_{k_1}(\mathbf{x}_1) w_{k_2}(\mathbf{x}_2) \int d\alpha_{k_1} d\alpha_{k_2} P(\alpha_{k_1}, \alpha_{k_2}) \alpha^*_{k_1} \alpha_{k_2}, \qquad (12)$$

$$G^{(2)}(\mathbf{x}_1, \mathbf{x}_2, \mathbf{x}_3, \mathbf{x}_4) = Tr(\rho E^{(-)}(\mathbf{x}_1) E^{(-)}(\mathbf{x}_2) E^{(+)}(\mathbf{x}_3) E^{(+)}(\mathbf{x}_4)) = \sum_{k_1, k_2, k_3, k_4} w^*_{k_1}(\mathbf{x}_1)$$
$$\times w^*_{k_2}(\mathbf{x}_2) w_{k_3}(\mathbf{x}_3) w_{k_4}(\mathbf{x}_4) \int d\alpha_{k_1} d\alpha_{k_2} d\alpha_{k_3} d\alpha_{k_4} P(\alpha_{k_1}, \alpha_{k_2}, \alpha_{k_3}, \alpha_{k_4}) \alpha^*_{k_1} \alpha^*_{k_2} \alpha_{k_3} \alpha_{k_4}, \qquad (13)$$

where $w_k(\mathbf{r},t) = v_k(\mathbf{r}) \exp(-i\omega_k t)$, $v_k(\mathbf{r}) = (\hbar \omega_k / 2)^{1/2} u_k(\mathbf{r})$, $\rho = \int P(\{\alpha\}) |\{\alpha\}\rangle\langle\{\alpha\}| d\{\alpha\}$ is the Glauber-Sudarshan P-representation of the density operator in terms of the projections on the Glauber coherent states and $P(\alpha_{k_1}, ..., \alpha_{k_n}) = \int P(\{\alpha\}) \prod_{\{\alpha\}\setminus\{\alpha_{k_1},...,\alpha_{k_n}\}} d\{\alpha\}$ are the marginal quasi-probability densities. The time-averaged space-integrated variance of the field's intensity is then equal to

$$(\Delta I_G)^2 \equiv \lim_{T\to\infty} \frac{1}{T} \int_{-T/2}^{T/2} dt \int d\mathbf{r} \{ G^{(2)}(\mathbf{r},t,\mathbf{r},t,\mathbf{r},t,\mathbf{r},t) - [G^{(1)}(\mathbf{r},t,\mathbf{r},t)]^2 \}. \qquad (14)$$

Substituting equations (12)-(13) into equation (14), we obtain:

$$(\Delta I_G)^2 = \int d\mathbf{r} \sum_{\substack{k_1,k_2,k_3,k_4 \\ k_1+k_2=k_3+k_4}} v^*_{k_1}(\mathbf{r}) v^*_{k_2}(\mathbf{r}) v_{k_3}(\mathbf{r}) v_{k_4}(\mathbf{r}) \int d\alpha_{k_1} d\alpha_{k_2} d\alpha_{k_3} d\alpha_{k_4}$$
$$\times \alpha^*_{k_1} \alpha^*_{k_2} \alpha_{k_3} \alpha_{k_4} [P(\alpha_{k_1},\alpha_{k_2},\alpha_{k_3},\alpha_{k_4}) - P(\alpha_{k_1},\alpha_{k_3}) P(\alpha_{k_2},\alpha_{k_4})]. \qquad (15)$$

In particular, for a state in which only a single mode, $E_k(\mathbf{r},t)$, is excited we have:

$$(\Delta I_G)^2 = (\Delta n_k)^2 (\hbar \omega_k / 2)^2 \int d\mathbf{r} |u_k(\mathbf{r})|^4, \qquad (16)$$

where

$$(\Delta n_k)^2 = \int d\alpha_k P(\alpha_k) |\alpha_k|^4 - \left( \int d\alpha_k P(\alpha_k) |\alpha_k|^2 \right)^2 = \overline{n_k^2} - (\overline{n}_k)^2 \qquad (17)$$

is the variance of the number of photons in the mode. Depending on the quasi-probability density distribution $P(\alpha)$, the value of $(\Delta n_k)^2$ in equation (6) can be can be close to $\overline{n_k^2}$ or to zero. In particular, for a pure coherent state, we have $P(\alpha) = \delta(\alpha - \alpha_0)$ and hence $(\Delta n_k)^2 = 0$. Obviously, this is a non-physical result, as the intensity measurement cannot have zero variance even for a pure coherent state [20]. This "paradox" appears because the above approach to calculation of field correlations neglects the commutators of the photon annihilation and creation operators which arise during the process of reduction of the expressions for correlation functions to a normally ordered form. The approximation usually works well for states with large numbers of photons, but it can be inaccurate when the number of photons is low. In particular, the contribution of vacuum fluctuations is ignored as a consequence of this approximation, which explains the difference between equation (16)-(17) and the exact results for the Fock and coherent states presented in equation (2)-(3) above. Note, however, that the consequences of neglecting the commutator $[a, a^\dagger]$ are not limited to the omission of vacuum fluctuations. Indeed, if one calculates the variance of the photon number operator $\hat{n} = a^\dagger a$ in a coherent state $|\alpha>$ using normal ordering and ignoring the commutators, i.e. approximating $(a^\dagger a)^2$ with $(a^\dagger)^2 a^2$ in the first term on the right-hand side of the expression $(\Delta n_k)^2 = <\alpha|(a^\dagger a)^2|\alpha> - (<\alpha|a^\dagger a|\alpha>)^2$, one would get $(\Delta n_k)^2 = 0$, instead of the correct answer $(\Delta n_k)^2 = |\alpha|^2$. This is consistent with the above "paradox" appearing in equations (16)-(17), and explains why we had to use direct calculations of the field's energy and its variance, instead of relying on generic quantum optical correlation functions.

In our experimental test, we have collected 1024 images of an unobstructed nearly-parallel wide monochromatic X-ray beam with energy of 30 keV using a Hamamatsu CMOS Flat Panel detector C9252DK-14 in "partial field" mode with pixel size 100 microns. We subsequently selected a uniformly illuminated region with 8 (vertical) × 256 (horizontal) pixels inside which we measured the average intensity and its variance. By binning pixels in the horizontal direction by the factors $2^n$, $n = 0,1,..5$, we were able to systematically vary the effective pixel size and, hence, the spatial resolution in the images. By adding image frames together into bunches of 1, 2, 4, 8 and 16 frames, we were able to independently vary the number of photons in each pixel. As the actual number of registered photons was unknown (because of the difficulties in measuring the detective quantum efficiency of the detector with sufficient accuracy), we present the output data as a function of the relative number of photons, $\bar{n}/\bar{n}_0$, where $\bar{n}_0$ is an unknown constant equal to the average total number of photons in a single image frame. Figure 1 depicts the obtained dependence of the quantity

$$\bar{n}_0 Q_2^2(M, \bar{n}) = \frac{M \, SNR^2 \, \bar{n}_0}{2\pi(\bar{n}+1/2)} = \frac{L^2}{2\pi[\Delta r(M)]^2} \frac{\tilde{I}_\psi^2}{(\Delta \tilde{I}_\psi)^2} \frac{\bar{n}_0}{(\bar{n}+1/2)}$$ on spatial resolution (number of

effective pixels $M$) and on the average total number of photons, $\bar{n}$.

**References**

[1] Heisenberg, W. Uber den anschaulichen Inhalt der quantentheoretischen Kinematik und Mechanik. *Z. Phys.* **43**, 172-198 (1927).

[2] Feynman, R. P., Leighton, R. B. & Sands, M. *The Feynman Lectures on Physics, Vol. 1,* (Addisson-Wesley, 1964).

[3] Angelow, A. Evolution of Schrödinger uncertainty relation in quantum mechanics. arXiv: 0710.0670.

[4] Epanechnikov, V. A. Non-Parametric Estimation of a Multivariate Probability Density. *Theory Probab. Appl.* **14**, 153-158 (1969).

[5] Silverman, B. W. *Density Estimation for Statistics and Data Analysis* (Chapman & Hall/CRC, 1998).

[6] de Hoog, F., Schmalz, G. & Gureyev, T. E. An Uncertainty Inequality. *Appl. Math. Lett.* **38**, 84-86 (2014).

[7] Mandel, L. & Wolf, E. *Optical Coherence and Quantum Optics* (Cambridge University Press, 1995).

[8] Robertson, H. P. The Uncertainty Principle. *Phys. Rev.* **34**, 163-164 (1929).

[9] Schrödinger, E. Sitzungsberichte der Preussischen Akademie der Wissenschaften, *Physikalisch-mathematische Klasse* **14**, 296-303 (1930).

[10] Braunstein, S. L., Caves, C. M. & Milburn, G. J. Generalized Uncertainty Relations: Theory, Examples, and Lorentz Invariance. *Annal. Phys.* **247**, 135-173 (1996).

[11] Folland, G. B. & Sitaram, A. The uncertainty principle: a mathematical survey. *J. Four. Anal. Applic.* **3**, 207-238 (1997).

[12] Ozawa, M. Universally valid reformulation of the Heisenberg uncertainty principle on noise and disturbance in measurement. *Phys. Rev. A* **67,** 42105 (2003).

[13] Giovannetti, V., Lloyd, S. & Maccone, L. Advances in quantum metrology. *Nature Photon.* **5**, 222-229 (2011).

[14] Xiao, Y. & Jing, N. Mutually Exclusive Uncertainty Relations. *Sci. Rep.* **6**, 36616 (2016).

[15] Faizal, M. Supersymmetry breaking as a new source for the generalized uncertainty principle. *Phys. Lett. B* **757,** 244-246 (2016).

[16] Barrett, H. H. & Myers, K. J. *Foundations of Image Science* (John Wiley & Sons, 2004).

[17] Gureyev, T.E. *et al.* Duality between noise and spatial resolution in linear systems. *Opt. Express* **22**, 9087-9094 (2014).

[18] Bach, R., Pope, D., Liou, S.-H. & Batelaan, H. Controlled double-slit electron diffraction. *New J. Phys.* **15**, 033018 (2013).

[19] Nesterets, Ya. I. & Gureyev, T. E. Young's double-slit experiment: noise-resolution duality. *Opt. Express* **23**, 3373-3381 (2015).

[20] Glauber, R. J. Optical Coherence and Photon Statistics in *Quantum Optics and Electronics* (ed. de Witt, C., Blandin, A. & Cohen-Tannoudji, C.) 61-185 (Gordon & Breach, 1965).

[21] Gureyev, T. E., de Hoog, F., Nesterets, Ya. I. & Paganin, D. M. On the noise-resolution duality, Heisenberg uncertainty and Shannon's information. *ANZIAM J.* **56**, C1-C5 (2015).

[22] Riek, C. *et al.* Direct sampling of electric-field vacuum fluctuations. *Science* **350**, 420-423 (2015).

**Acknowledgements**

T.E.G. thanks A. Yu. Ignatiev and F. de Hoog for useful discussions relevant to the early development of this paper.

**Author contributions**

T.E.G. conceived the main results, participated in the X-ray experiment and led the writing of the paper. A.K. carried out most of the analytical calculations and participated in development of key aspects of the study. Y.I.N. led the X-ray experiment and analysed the experimental data. H.M.Q. and D.M.P. provided multiple theoretical ideas implemented in the paper, participated in development of the research and in writing of the paper.

**Additional Information**

Competing financial interests: The authors declare no competing financial interests.

**Figure legends**

Figure 1. Measured dependence of the quantity $Q_2^2$ on the spatial resolution (number of effective pixels *M*) and on the total number of photons.